\documentclass[
 reprint,
superscriptaddress,
 amsmath,amssymb,
 aps,
pra,
]{revtex4-1}
\usepackage[ colorlinks = true,
            linkcolor = blue,
              urlcolor  = blue,
              citecolor = red,
              anchorcolor = green,
 ]{hyperref}
\usepackage{graphicx}
\usepackage{color}
\usepackage{dcolumn}
\usepackage{bm}
\usepackage{xfrac}

\newcommand\eq[1]{Eq.~(\ref{eq:#1})}

\newcommand\fig[1]{Fig.\ref{fig:#1}}

\newcommand\bra[1]{\left\langle\,#1\,\right|} 
\newcommand\ket[1]{\left|\,#1\,\right\rangle}

\begin{document}

\preprint{APS/123-QED}

\title{Generalized overlap quantum state tomography}
\email{Contribution of NIST, an agency of the U.S. government, not subject to copyright.}
\author{Rajveer Nehra}
\email{rn2hs@virginia.edu}
 \affiliation{Department of Physics, University of Virginia, 382 McCormick Rd, Charlottesville, VA 22904-4714, USA}
\author{Miller Eaton}
\email{me3nq@virginia.edu}
\affiliation{Department of Physics, University of Virginia, 382 McCormick Rd, Charlottesville, VA 22904-4714, USA}
\author{Carlos Gonz\'alez-Arciniegas}
 \affiliation{Department of Physics, University of Virginia, 382 McCormick Rd, Charlottesville, VA 22904-4714, USA}
\author{M. S. Kim}
\email{m.kim@imperial.ac.uk}
\affiliation{QOLS, Blackett Laboratory, Imperial College London, SW7 2AZ, UK}
\affiliation{Korea Institute of Advanced Studies, Seoul, 02455, Korea}
\author{Thomas Gerrits}
\affiliation{National Institute of Standards and Technology, 100 Bureau Drive, Gaithersburg, MD 20899, USA}
\author{Adriana Lita}
\affiliation{National Institute of Standards and Technology, 325 Broadway, Boulder, CO 80303, USA}
\author{Sae Woo Nam}
\affiliation{National Institute of Standards and Technology, 325 Broadway, Boulder, CO 80303, USA}
\author{Olivier Pfister}
\email{olivier.pfister@gmail.com}
\affiliation{Department of Physics, University of Virginia, 382 McCormick Rd, Charlottesville, VA 22904-4714, USA}

\date{\today}
\begin{abstract}
We propose and experimentally demonstrate a quantum state tomography protocol that generalizes the Wallentowitz-Vogel-Banaszek-W\'odkiewicz point-by-point Wigner function reconstruction. The full density operator of an arbitrary quantum state is efficiently reconstructed in the Fock basis, using semidefinite programming, after interference with a small set of calibrated coherent states. This new protocol is resource- and computationally efficient, is robust against noise, does not rely on approximate state displacements, and ensures the physicality of results.
\end{abstract}

 \maketitle

\textit{Introduction}.   Since a quantum system is fully characterized by its density operator~\cite{Fano1957}, the experimental implementation, and investigation, of quantum state tomography~\cite{Paris2004} plays a crucial role in quantum information (QI). While the dimension $2^{N}$ of a $N$-qubit Hilbert space prohibits full quantum state tomography for large values of $N$, except in the particular case of sparse density operators~\cite{Cramer2010},  full state tomography of single, or few, quantum systems can still be realized and is  essential to characterizing important resource states, e.g.\ quantum error correcting codes. 
Here, we focus on bosonic quantum modes, a.k.a.\ qumodes, as implemented in general by vibrational eigenmodes of quantum harmonic oscillators and, in particular, by quantum electromagnetic fields as used in continuous-variable (CV) QI~\cite{Furusawa2011,Pfister2019}, the latter being particularly well suited to the generation of massively scalable multipartite entanglement, in particular of the universal quantum computing substrates that are cluster states~\cite{Pysher2011,Chen2014,Roslund2014,Yokoyama2013,Yoshikawa2016,Asavanant2019,Larsen2019}.  

In the CVQI context, the Wigner function~\cite{Wigner1932,Hillery1984} plays a central role as a quantum state descriptor strictly equivalent to the density operator $\rho$:
\begin{equation}
W(q,p) =  \frac{1}{\pi}\int_{-\infty}^{\infty}e^{2ipy}\langle{q- y}|{\rho}|{q+ y}\rangle dy,
\end{equation}
where quantum phase space variables $q$ and $p$ are the eigenvalues of the position-like amplitude quadrature, $\hat{Q}=(\hat{a}+\hat{a}^\dag)/\sqrt2$, and momentum-like phase quadrature, $\hat{P}=i(\hat{a}^\dag-\hat{a})/\sqrt2$, of the electromagnetic field, and where $\hat{a}$ is the boson annihilation operator for a given qumode, typically specified by its wave vector and polarization. The experimental determination of the Wigner function, first proposed and realized by way of interferometric, homodyne quadrature measurements~\cite{Smithey1993}, thus constitutes another approach to quantum state tomography. A technical difficulty of the aforementioned optical homodyne tomography approach resides in the need for computationally intensive reconstruction procedures, using either the inverse Radon transform (whence the ``tomography'' moniker) or maximum likelihood algorithms~\cite{Lvovsky2009}.
\\
\indent Such difficulties can be alleviated by replacing field measurements with photon-number ones, using the fact that the Wigner function at the origin of phase space co\"incides with the expectation value of the photon-number parity operator~\cite{Royer1977}. This yields an expression of the Wigner function in the Fock basis which is easy to reconstruct, as was first proposed by Wallentowitz-Vogel~\cite{Wallentowitz1996} and Banaszek-Wodkiewicz~\cite{Banaszek1996} (WVBW). In this method, a simple phase-space translation, i.e., displacement, of the quantum state to be characterized, followed by parity measurements, allows easy determination of the Wigner function. This was first implemented experimentally on the phononic field of vibration of a single trapped ion~\cite{Leibfried1996}, as well as on microwave cavity fields~\cite{Bertet2002,Vlastakis2013}. More recently, the coming of age of photon-number-resolving (PNR) detection~\cite{Lita2008} has opened the door to using the full WVBW method on traveling optical fields with no prior knowledge of the measured quantum state~\cite{Sridhar2014a,Nehra2019}.  
While the WVBW method presents clear advantages in terms of the numerical demands on reconstruction, it requires a phase space raster scan involving a large number of optical displacements, and the pitch of the raster scan is determined by the specific features of the---unknown---Wigner function to be resolved. Moreover, the best experimental implementation of phase space displacements is intrinsically lossy~\cite{Paris1996}. 
Finally, the method does require, like homodyne tomography, a very high system detection efficiency in order to prevent the quantum decoherence caused by vacuum fluctuation contamination.
Additionally, the WVBW protocol mandates a matrix inversion for each experimental data point in order to infer the true photon-number distribution from the measured loss-degraded distribution which could be experimentally demanding for probing the Wigner functions of complicated structure, such as cat states or Gottesman-Kitaev-Preskill (GKP) states~\cite{Laiho2010}.

\indent In this paper, we present a generalization of the WVBW approach which uses a Wigner function overlap measurement to reconstruct the density operator, rather than the Wigner function, using computationally efficient semidefinite programming. This general method requires considerably less data acquisition, and ensures physical results which are robust against  measurement noise. The effect of known system losses can also be entirely deconvoluted from the measured density operator. We present the mathematical derivation of this generalized overlap quantum state tomography and present experimental results for a single-photon Fock state with performance that far exceeds that of the recent WVBW demonstration~\cite{Nehra2019}.  Furthermore, we can perform loss-compensation in one fell swoop for the entire density matrix $\rho$, unlike at each experimental data point in WVBW method.

\textit{Theory}.  We consider the situation depicted in \fig{schematic}(a): 
\begin{figure}[h]
    \centering
\includegraphics[width = 0.485\textwidth]{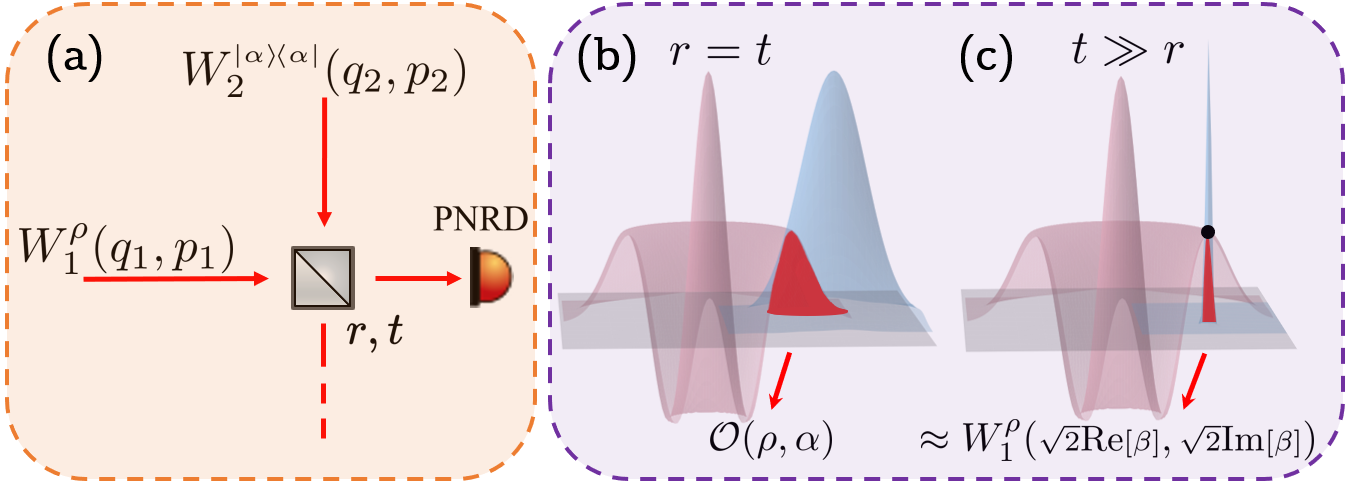}
\caption{(a), Schematic of the experiment: the field to be measured, of density operator $\rho$, interferes with a calibrated field in coherent state $\alpha$ at a beamsplitter of field reflectance $r\in\mathbb R$ and transmittance $t=(1-r^{2})^{1/2}$. PNRD: photon-number-resolving detector. (b), Principle of generalized overlap tomography exemplified with a two-photon Fock state. (c), Limit case of (b), where a highly unbalanced beamsplitter merely implements a  displacement of $\rho$ by  $-\beta$.} 
    \label{fig:schematic}
\end{figure}
a field with unknown density operator $\rho$ and Wigner function $W_1(q_1,p_1)$ 
interferes with a reference field in a coherent state $\ket\alpha\bra\alpha$ of Wigner function $W_2(q_2,p_2)$. We then simply count photon statistics at only one beamsplitter output using the PNR detector; using these, we evaluate the expectation value of the photon number parity operator, i.e., the value of the origin of the Wigner function of this output mode~\cite{Kim2002wig_overlap,SupMat}
\begin{equation} 
    W'_{1}(0,0;r,t)=\frac{1}{r^2}\iint W_{\rho}(q,p)\,W_{|\alpha\rangle\langle \alpha|}(\tfrac trq,\tfrac trp)\,dq\,dp,
     \label{eq:wig_overlap}
\end{equation}
as illustrated in \fig{schematic}(b). Setting $r=t$ in the above formula yields, by virtue of the Wigner function overlap theorem~\cite{Leonhardt1997}, the overlap $\mathcal{O}$ of the unknown  $\rho$ with $|\alpha\rangle \langle \alpha|$:
\begin{equation}
   \mathcal{O}=\text{Tr}[\rho|\alpha\rangle\langle\alpha|] = \pi W'_{1}(0,0;\tfrac1{\sqrt2},\tfrac1{\sqrt2}).
    \label{eq:wig_overlap_balance}
\end{equation}
Note that $\mathcal{O}$ is proportional to the Husimi Q function, $Q(\alpha)$, which we sample sparsely~\cite{Husimi}. Note also that,
in the limit case $t$$\gg$$r$,  the function $W_{2}(\tfrac trq,\tfrac trp)$ in \eq{wig_overlap} is a contracted Gaussian that tends toward a Dirac delta function $\delta(\sqrt2\text{Re}[\beta],\sqrt2\text{Im}[\beta])$, where $\beta = r\alpha/t$,  
thereby yielding $W_{1}(\sqrt2\text{Re}[\beta],\sqrt2\text{Im}[\beta])$, i.e., precisely the WVBW tomography protocol, as illustrated in \fig{schematic}(c). The validity of this limit case is equivalent to the validity of implementing a displacement with an unbalanced beamsplitter. Our new overlap approach is free of such considerations.

From here on, we set $r=t$ for simplicity. We provide a treatment of the general case of arbitrary $r,t$ in the supplemental document~\cite{SupMat}. 
Even though this would appear to cause an irremediable loss of information (the role of the other output port is examined in the supplemental document~\cite{SupMat}), we show that $\rho$ can nonetheless be accurately and efficiently retrieved by measuring $\mathcal{O}_{j}$ for a series of distinct coherent states $|\alpha_j\rangle$. In the Fock basis, we get
\begin{equation}\label{eq:overlap}
    \mathcal{O}_j=\sum_{n=0}^{n_0}\sum_{m=0}^{n_0}c_{jn}^{*}c_{jm}\rho_{nm},
\end{equation}
where $c_{jm}$=$\exp(-|\alpha_j|^2/2)\alpha_j^m/\sqrt{m!}$ and the size of the Hilbert space is $n_0+1$. Our superconducting transition-edge sensor (TES) has high-efficiency PNR capabilities up to 5 photons at 1064\,nm, leading to our choice of $n_0=5$~\cite{Lita2008,Sridhar2014a}.

For $N=(n_0+1)^2$ measurements, we can write \eq{overlap} in matrix form in $N$-dimensional  Liouville space
\begin{equation}\label{eq:mat}
\mathbf O=\mathbf C\mathbf P,
\end{equation}
where $\mathbf{O}$=$(\mathcal O_{j})_{j}^T$,  $\mathbf C$=$(c_{jn}^{*}c_{jm})_{j,nm}$, and $\mathbf P$=$(\rho_{nm})_{nm}^{T}$\cite{SupMat}. Inverting \eq{mat} yields the unknown  Liouville vector $\mathbf P$. To do this, we employ semidefinite programming (SDP) to run a convex quadratic optimization algorithm that minimizes the $\ell^2$-norm, $||\mathbf{O-CP}||_2$, subject to physicality constrains in order to extract $\mathbf{P}$. The procedure is computationally efficient and yields a unique solution~\cite{boyd2004convex}. Note that $\bf C$ does not have to be a square matrix, so that the number of measured overlaps (the dimension of $\mathbf{O}$) can be increased for better data statistics.

A crucial point is the impact of inevitable experimental fluctuations on  the numerical stability of the solution. Indeed, by nature of its slowly-decaying Poissonian coefficients, matrix $\bf{C}$ necessarily contains both large and small entries and, therefore, both large and small singular values, which make it ill conditioned~\cite{Tarantola2005} and make its inversion extremely sensitive to experimental fluctuations in the measured photon statistics. In order to suppress these instabilities, we choose to use a Tikhonov regularization procedure~\cite{Tikhonov1943}, formulated as the SDP problem
\begin{align}
&\text{Minimize}\quad||{\bf{\mathbf{O}-CP}}||_2 + \gamma{\bf{||P||_2}}\nonumber\\
&\text{Subject to}\quad\rho\geq0, \quad \text{Tr}[\rho] = 1,
\end{align}
where $\gamma$ is a small regularization parameter set according to the noise level~\cite{Flammia2012} and the optimization remains quadratic convex. 

Another crucial point is the effect of decoherence on density matrix reconstruction. Optical losses to the state before and after interference with $|\alpha_j\rangle$, including detector efficiency, can be modeled by inserting a single fictitious beamsplitter of transmittivity $\eta$ in front of a perfect detector 
that both leaks light out and couples in vacuum fluctuations as discussed in detail in the supplemental document~\cite{SupMat}. This leads to the resulting, binomial-law density matrix,  

\begin{equation}
    \rho'_{nm}=\sum_{k =0}^{\infty}\rho_{n+k, m+k} 
    {\textstyle{\binom{n+k}{k}}^{\frac{1}{2}}{\binom{m+k}{k}}^{\frac12}}(1-\eta)^{k}\eta^{\frac{n+m}{2}},
    \label{eq:rho_loss}
\end{equation}
which can be inverted to recover $\rho$ from $\rho'$~\cite{Kiss1995}. If we only consider the diagonal elements of $\rho$, then this inversion is exactly described by correcting a loss-degraded photon-number distribution, 
as has been experimentally demonstrated for state characterization~\cite{Achilles2006} and the WVBW protocol, but  requiring  a matrix inversion for each experimental data point~\cite{Laiho2010}.  In our general case, we can perform such an inversion in one fell swoop for the whole $\rho$, in lieu of entry-wise as above. The same difficulty arises, though, of high sensitivity of the inversion procedure to small experimental fluctuations in  $\rho'$ which can lead to unphysically large or negative diagonal density matrix elements in the reconstructed $\rho$~\cite{SupMat}.  
Again, we solve this problem by  SDP:
\begin{align}
&\text{Minimize}\quad\sum_{i=0}^{N_{max}}|| \rho'^{(i)}
-\mathbf M^{(i)}(\eta)\rho^{(i)}
||_2\\
&\text{Subject to}\quad\rho\geq0, \hspace{2mm} \text{Tr}[\rho] = 1, \text{ and} \hspace{2mm} \rho_{nn}\leq \eta^{-n}\rho'_{nn},\nonumber
\label{eq:minimize_loss}
\end{align}
where $\rho^{(i)}
$ denotes the $i^\text{th}$ diagonal of $\rho$ ($i=0$: main diagonal) and $\mathbf M^{(i)}(\eta)$ is the matrix describing the binomial-law loss degradation along each diagonal of $\rho$. The third constraint stems from the fact that \eq{rho_loss} yields $\rho'_{nn}=\eta^n\rho_{nn}+\epsilon$, where $\epsilon$ is positive. If the value of the loss parameter $\eta$ is known, this loss deconvolution method is very efficient and reliable, as is further detailed in the supplemental material where we provide a side-by-side comparison of the improvement over an analytical approach in the presence of noise~\cite{SupMat}. 

The case of mode mismatch deserves a separate mention. In contrast to homodyne detection, nonideal-contrast interference between the coherent-state and signal fields can't simply be treated as loss.  To account for mode mismatch, we consider a multimode detection theory where the coherent state is decomposed into $|\sqrt{\mathcal{M}}\alpha^j\rangle_{\parallel}$, which interferes entirely with the signal, and an orthogonal component, $|\sqrt{1-\mathcal{M}}\alpha^j\rangle_{\perp}$, which interferes with vacuum~\cite{tichy2014interference}. Parameter $\mathcal{M}$ is determined by the degree of overlap, and can be calculated from a bright-field visibility measurement~\cite{Banaszek2002}. Because the PNR detectors used herein are mode insensitive, the total measured signal is represented by the sum of detected modes. 
This yields a measured photon number distribution that is a convolution of the individual mode distributions~\cite{MandelWolf}. We measured a mode overlap parameter of $\mathcal{M}=0.83(2)$ (0.86(2)) when performing the tomography of the coherent (Fock) state and deconvolve the Poissonian distribution of the mode-mismatched $|\sqrt{1-\mathcal{M}}\alpha^j\rangle_{\perp}$ from our measurements. It is important to note that the $\mathcal{O}$ values obtained from the expectation of parity are now between $\rho$ and $|\sqrt{\mathcal{M}}\alpha^j\rangle$ for each coherent state probe, and therefore the coefficient matrix $\mathbf{C}$ must be modified accordingly as further detailed in the supplemental material~\cite{SupMat}. 

\textit{Experimental implementation}. The experimental setup was identical to our previous implementation of WVBW tomography~\cite{Nehra2019} and is described in detail in the supplemental material~\cite{SupMat}. It was based on a very stable CW Nd:YAG laser which provided all coherent states upon phase and amplitude modulations by a piezoelectric-actuated mirror and a home-made $\rm RbTiOAsO_{4}$ electro-optic modulator, respectively. The calibrated coherent-state amplitude range was $|\alpha|=0.138(2)$ to $0.339(3)$, in six steps, directly calibrated using a TES. The coherent-state probe amplitudes were calibrated by comparing the TES photon statistics to that of a Poisson distribution with the signal beam blocked as detailed in our previous implementation~\cite{Nehra2019}.
The phase scan was ten discrete steps of $0.58(5)$ radians each. The laser was also resonantly frequency-doubled to pump an optical parametric oscillator whose narrowband pair emission provided heralded single-photon Fock states~\cite{Nehra2019}. All data acquisition was computer-controlled.

\begin{figure}[!hbt]
    \centering
    \includegraphics[width = 0.9\columnwidth]{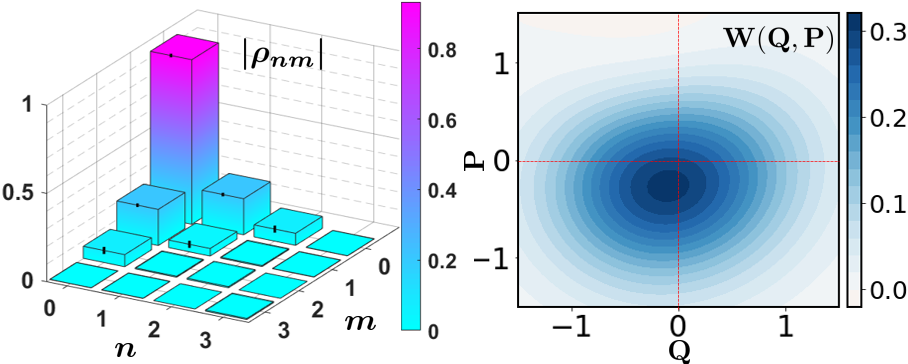}
    \caption{Left, absolute value of the experimentally reconstructed density matrix elements for the coherent state. Right, associated Wigner function. The black error bars are obtained from the measurement statistics.}
    \label{fig:coher_results}
\end{figure}

\textit{Results. Coherent state}. We implemented the generalized overlap tomography protocol for a weak coherent state and a single-photon state. The rationale for measuring a coherent state was to display a phase-dependent, i.e. non-cylindrically symmetric structure in phase space. The coherent state $|\beta\rangle$ was chosen $|\beta|=0.191(3)$, as calibrated by the TES Poissonian photon statistics. For each of the 60 coherent-state probes $|\alpha_{j}\rangle$, data was acquired for 3 s to obtain $\sim 10^5$ events from which to construct the photon-number probability distributions. The SDP tomography results after correcting for mode mismatch are displayed on \fig{coher_results}.
\begin{figure*}[!htb]
\centerline{\includegraphics[width= 1.7\columnwidth]{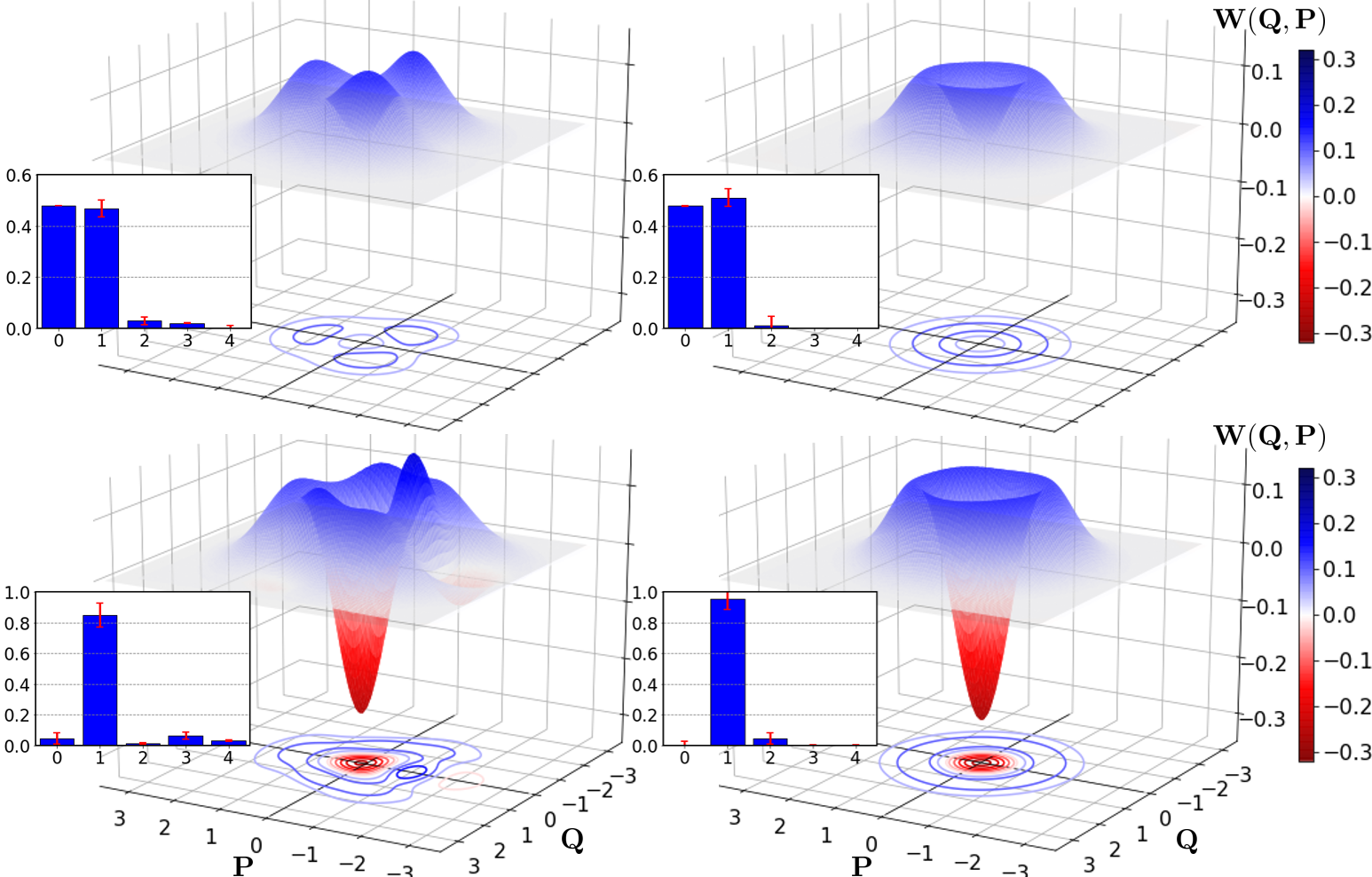}}
\caption{ 
Generalized overlap tomography of a single-photon Fock state. Top row, SDP with mode mismatch corrected but no correction for losses. Bottom row, loss-deconvoluting SDP reconstruction. Left column, direct reconstructions. Right column, reconstructions using phase-averaged measurements. Reconstruction fidelities are 0.85(8) for the bottom left panel and 0.94(6) for the bottom right panel. Inset: photon-number probability distribution. }
    \label{fig:fock_results}
\end{figure*}
 
Examining the magnitude of the density matrix elements, we clearly see that the diagonal and off-diagonal terms of $\rho$ were both successfully reconstructed. The phase and amplitude accuracy is more evident when comparing the associated Wigner functions plotted in \fig{coher_results}, right, where the red dashed lines delineate the zero axis values. 
We achieve a fidelity of $\mathcal{F}=0.97(2)$ between the reconstructed state, $\rho$, and the target pure state, $|\beta\rangle$. The slight asymmetry of the Wigner function is imputable to residual phase noise in our measurements (only passive noise cancellation techniques were used for the optical paths).

\textit{Single-photon Fock state}. The Wigner functions obtained from the reconstructed density matrix are shown in \fig{fock_results}. Due to the nature of our heralded source undergoing an overall loss, $\eta$, we expect to measure a statistical mixture of the one-photon and vacuum states which has a rotationally symmetric Wigner function~\cite{Lvovsky2001,Nehra2019}. Under this assumption, an average over the optical phases of the coherent probes can be performed, yielding the results on the right column of \fig{fock_results}. It is, of course, also interesting to examine the unaveraged measurements, left column of \fig{fock_results}, in order to assess the quality of our tomographic reconstruction. While the effect of experimental phase noise is, again, visible, the analytic form of the function is well defined. Finally, the performance of the noise deconvolution by SDP is displayed on the bottom row of \fig{fock_results}. The overall loss was determined to be $\eta=0.50(1)$ by measuring the heralding ratio, as was done in Ref.~\citenum{Nehra2019}. Assuming no prior knowledge about the state other than this calibrated measurement loss, the reconstructed loss-compensated state is depicted in the bottom left of Fig.~\ref{fig:fock_results}, where we achieved a fidelity of $\mathcal{F}=0.85(8)$ with a single-photon Fock state.  
Adding the assumption of a phase-invariant state and averaging measurements  for each of the ten phases before compensating for loss yielded the nearly perfect reconstruction shown in the bottom right panel, where we achieved  $\mathcal{F}=0.94(6)$. It is worth emphasizing that the negativity of the single-photon Wigner function was fully recovered after compensating for loss and mode-mismatch 
(\fig{fock_results}, bottom row), even though the  50\% loss level suppressed negativity when no loss deconvolution was performed (\fig{fock_results}, top row). 
Finally, it is important to note that the maximum amplitude probe was  $|\alpha_\text{max}|\simeq0.34$, which led to a mean photon number detection of $\langle N\rangle\simeq0.56$, yet our overlap tomography accurately reconstructed the Wigner function at quadrature coordinates beyond $q\text{ or }p=3$ (consistent with our truncation of the Hilbert space to $n_{o}$=5). This is in stark contrast to the WVBW case of Ref.~\citenum{Nehra2019}, in which the maximum of the Wigner function, at $q\text{ or }p=1$, could not be reached using displacements with $|\alpha_\text{max}|\simeq0.80$ and $\langle N\rangle\simeq1.64$. Therefore, generalized overlap tomography necessitates PNR detection of significantly lower photon flux while still requiring the detection of only a single field mode. \\
\textit{Conclusion}. We proposed and experimentally demonstrated generalized overlap quantum state tomography using PNR measurements on a single field-mode. Our approach, {\em (i)}, makes no prior assumption on the initial state {\em (ii)}, exploits numerically efficient, noise-robust SDP that enforces physicality, {\em (iii)}, uses fewer, lower-amplitude probes than point-by-point WVBW tomography (thus likely outperforming 
WVBW for complex states where fine resolution of the Wigner function is required), {\em (iv)},  implements no approximated displacement operations, {\em (v)}, requires only a \textit{single} PNR detector and necessitates fewer measurements than densely probing the Wigner function~\cite{Shen2016}, {\em (vi)}, compensates for known losses with fewer numerical instabilities.

Our approach is equally valid for other physical systems and can be readily applied in circuit quantum electrodynamics~\cite{Gao2018}. It could also be used to directly measure the purity of a quantum state by measuring the overlap between two copies of the same system, which allows access to the second order R\'enyi entropy extensively used to quantify the entanglement of many-body physical systems~\cite{Islam2015}. Finally, the proposed scheme can be simply extended to characterize a multi-mode quantum system by interfering with a multi-mode set of coherent states followed by measuring the overall parity of the state after the interference.

\textit{Acknowledgements}. The work was supported by NSF grants PHY-1708023 and PHY-1820882, by the KIST OPEN Research Programme, and by a KIAS visiting professorship. RN, ME, CGA, and OP thank Nicolas Fabre and Paulo A. Nussenzveig  for helpful discussions.  MSK thanks the Royal Society, Samsung GRO grant and the EPSRC (EP/R044082/1). RN thanks Liang Jiang, Joshua Combes, and Jonathan Gross for fruitful discussions in the 2019 Byron Bay Quantum Invitational.\\
${\dagger, \ddagger}$ contributed equally to this work.

\bibliography{Pfister,Pfister2}

\begin{widetext}
\section*{Supplemental material}
Here, we provide supplementary information for ``Generalized overlap quantum state tomography.''  In Section \ref{sec:1}, we provide the formalism for measuring the Wigner function overlap of quantum optical states. This formalism is extended in Section \ref{sec:gen_bs} to determine the precise form of the overlap matrix elements in the case of an arbitrary unbalanced beamsplitter. Section \ref{sec:tomo} discusses the tomographic reconstruction utilizing semidefinite programming (SDP), and Section \ref{sec:exp} details the experimental implementation. Section \ref{sec:loss_comp} details computationally efficient and robust loss compensation scheme using SDP, while Section \ref{sec:pn_dist_equiv} demonstrates the equivalence of loss to the unknown state before and after interference with coherent-state probes. Finally, Section \ref{sec:mode_mismatch} describes the effects of imperfect mode-matching between the signal and probe fields.
\section{General Wigner function overlap}\label{sec:1}

Sending an unknown state, $\rho_{in}$, and a probing coherent state, $|\alpha_j\rangle \langle \alpha_j|$, through a beamsplitter and measuring the Wigner function at the origin of one output mode directly yields the Wigner function overlap between $\rho_{in}$ and a probe. To see this, we adopt the Heisenberg picture and determine the evolved output quadratures under the beamsplitter interaction to be $q_1'= tq_1 - rq_2$ and  $p_1' =  tp_1 - rp_2$. Likewise, $q_2' =  rq_1 + tq_2$ and $p_2' = rp_1 + tp_2$.\\
The two-mode input state is written in the Wigner function representation as
\begin{equation}
    W_{1,2}({\bf{x}})= W_1(q_1, p_1)W_2(q_2, p_2). 
\end{equation}
Next, by using the evolved quadratures, one can write the Wigner function of the beamsplitter output as
\begin{align}
W'_{1,2}({\bf{x'}}) =  &W_1(tq_1'+rq_2', tp_1'+rp_2') W_2(-rq_1'+tq_2', -rp_1'+tp_2'),
\label{eq:2mode_Wig}
\end{align}
where ${\bf{x}}$ and ${\bf{x'}}$ are column vectors consisting of quadratures corresponding to the input and output modes, respectively. The value of the Wigner function of output mode 1 at the origin can be obtained by setting $q_1'=p_1'=0$ and tracing out over mode 2 to yield
\begin{align}
   \iint  W'_{1,2}({{\bf{x'}}})dq_2'dp_2'|_{q_1', p_1' = 0}
   &=\iint W_1(rq_2',rp_2')W_2(tq_2',tp_2')dq_2'dp_2'\\
   &=\frac{1}{r^2}\iint W_1(q,p)W_2(\tfrac{t}{r}q,\tfrac{t}{r}p)dqdp.
    \label{eq:meas_overlap}
\end{align}
Setting $r=t=\frac{1}{\sqrt{2}}$ yields the Wigner function overlap between $\rho_{in}$ and $|\alpha_j\rangle \langle \alpha_j|$, i.e., Eq.~3 in the main text. 
In the case of an unbalanced beamsplitter where the input probe is still a coherent state, $\alpha_j=\tfrac{1}{\sqrt{2}}(q_{\alpha_j}+ip_{\alpha_j})$, we now have

\begin{align}
    W_2(\tfrac{t}{r}q,\tfrac{t}{r}p)=W_{\ket{\alpha_j}\bra{\alpha_j}}(\tfrac{t}{r}q,\tfrac{t}{r}p)&=\frac{1}{\pi}\exp\Big[-\big(\tfrac{t}{r}q+q_{\alpha_j}\big)^2-\big(\tfrac{t}{r}p+p_{\alpha_j})\big)^2\Big]\\
    &=\frac{1}{\pi}\exp\left\{-\frac1{\sigma^2}\left[\left(q+q_{\beta_j}\right)^2+\left(p+ p_{\beta_j}\right)^2\right]\right\},
\end{align}
where $\sigma=\frac{r}{t}$ and $\beta_j=\sigma \alpha_j$. The integral overlap from Eq.~\ref{eq:meas_overlap} is then 

\begin{equation}
  \frac{1}{t^2}\iint W_1(q,p)W_{\sigma}(q+q_{\beta_j},p+p_{\beta_j})dqdp, 
    \label{eq:meas_overlap2}
\end{equation} 
where $W_{\sigma}(q+q_{\beta_j},p+p_{\beta_j}) = \frac{1}{\pi\sigma^2}\exp\left\{-\frac1{\sigma^2}\left[\left(q+q_{\beta_j}\right)^2+\left(p+ p_{\beta_j}\right)^2\right]\right\}$. When $\sigma>1$, Eq.~\eqref{eq:meas_overlap2} gives a scaled overlap between $\rho_{in}$ and a thermal state displaced by $\beta_j$. When $\sigma<1$, however, the overlap is between $\rho_{in}$ and a mathematical object that approaches a displaced delta function in the limit where $\sigma \rightarrow0$ and $|\alpha| \rightarrow \infty$, which exactly probes the Wigner function of $\rho_{in}$ point-by-point as in the unbalanced homodyne technique of Refs.~\cite{Wallentowitz1996} and~\cite{Banaszek1996}.

One might also wonder about the outcome of a similar measurement performed on the other port of the beamsplitter. If we go back to Eq.~\eqref{eq:2mode_Wig} and now determine the value of the Wigner function of output mode 2 at the origin while tracing out over mode 1, we have
\begin{align}
   \iint  W'_{1,2}({{\bf{x'}}})dq_1'dp_1'|_{q_2', p_2' = 0}
   &=\iint W_1(tq_1',tp_1')W_2(-rq_1',-rp_1')dq_1'dp_1'\\
   &=\frac{1}{t^2}\iint W_1(q,p)W_2(-\tfrac{r}{t}q,-\tfrac{r}{t}p)dqdp.
   \label{eq:meas_overlap_mode2}
\end{align} 
This time when we set $r=t$, the measured Wigner function overlap is between $\rho_{in}$ and $|-\alpha_j\rangle \langle -\alpha_j|$, i.e., the very same coherent state probe with a phase factor of $e^{i\pi}$. From this, we can conclude that measuring the Wigner function at the origin of both beamsplitter outputs  would yield the overlap between the signal and coherent-state probes at opposite phases. When performing the tomographic reconstruction, it is possible to utilize both outputs to collect overlap measurements and only externally vary the probe phases by half of the desired range; however, ensuring that both detection channels following the beamsplitter are identical in losses, detector efficiency, etc., is experimentally challenging, and this also imposes additional requirements on PNR detection capabilities.  Therefore, it is simpler to utilize a single output mode to perform the tomographic reconstruction and correct for known losses as detailed below and in the main text.

\section{Discussion on inverting Eq. 4 for a general beamsplitter}\label{sec:gen_bs}
In this section, we investigate an inversion scheme for an arbitrary beamsplitter with reflection and transmission coefficients of $r$ and $t$, respectively. We start with formally defining the Wigner function of an operator denoted by $\hat{T}$ as 

\begin{equation}
W_{\hat{T}}(q,p)=\frac{1}{2\pi} \text{Tr}[\hat{T} \hat{\Pi}(q, p)],
 \label{eq:Wig_general_def_1}
\end{equation}
where $\hat{\Pi}(q, p)$ is the translated parity operator formally defined as
\begin{align}
 \hat{\Pi}(q, p)=\iint \frac{d q^\prime d p^\prime}{2 \pi} e^{-i(qp^\prime-p q^{\prime})} \hat{D}\left(q^{\prime}, p^{\prime}\right)=\int dq^\prime e^{-ipq^\prime}\ket{q+\frac{q^\prime}{2}}\bra{q-\frac{q^\prime}{2}},
 \label{eq:Wig_general_def}
\end{align}	
where $\hat{D}(q,p)$ is the phase space displacement operator. 
For a given quantum state $\hat{T} = \rho$, Eq.~\ref{eq:Wig_general_def} leads to the usual Wigner function of the state.

However, this definition is general and may be extended to any arbitrary operator, $\hat{T}=T(\hat{q},\hat{p})$ in order to calculate the so called Weyl symbol representing the operator $\hat{T}$. This is achieved by inverting Eq.~\ref{eq:Wig_general_def_1} which results to the operator $\hat{T}$ in Weyl symbol form as

\begin{align}
	\hat{T}=\iint dpdq W_T(q, p) \hat{\Pi}(q, p).
	\label{eq:T_weyl}
\end{align}
Here we have used the fact that Tr$[\hat{\Pi}\left(q,p\right) \hat{\Pi}\left(q^{\prime}, p^{\prime}\right)]=2 \pi \delta(q-q^{\prime})\delta(p-p^{\prime})$.
 
Next, we calculate the matrix elements of the operator $\hat{T}$ as

\begin{align}\label{eqn:OpEle}
{T}_{n, m}=\bra{n}\hat{T}\ket{m}=\iint  W_T(q, p)\bra{n}\hat{\Pi}(q, p)\ket{m}dqdp,
\end{align}
where the matrix elements of the displaced parity operator can be determined using Eq.~\ref{eq:Wig_general_def} as
\begin{align}
\bra{n}\hat{\Pi}(p, q)\ket{m}&=\int dq^{\prime} e^{-i p q^{\prime}} \bigg{\langle} n\bigg{|}{q-\frac{q^{\prime}}{2}} \bigg{\rangle}\bigg{\langle} q+\frac{q^{\prime}}{2}\bigg{|}{ m}\bigg{\rangle}\nonumber \\
&=\frac{e^{-q^{2}}}{\sqrt{\pi 2^{m+n} n ! m !}} \int d q^{\prime} e^{-i p q^{\prime}} e^{-q^{\prime 2} / 4} H_{n}\left(q+\frac{q^{\prime}}{2}\right) H_{m}\left(q-\frac{q^{\prime}}{2}\right)\nonumber \\
&\overset{q^\prime\rightarrow 2(x-ip)}{=}\frac{2 e^{-\left(q^{2}+p^{2}\right)}}{\sqrt{\pi 2^{n+m} n ! m !}} \int d x e^{-x^{2}} H_{n}(x+(q-i p))(-1)^{m}H_m(x-(q+i p)).
\end{align}
Note that 
$\langle x | n\rangle=\frac{1}{\pi^{1/4}} \frac{e^{-x^{2} / 2}}{\sqrt{2^n n!}} H_{n}(x)$ and  $H_n(-x)=(-1)^nH_n(x)$.
Using these relations, we get
\begin{align}
\int d x e^{-x^{2}} H_{m}(x+\sigma) H_{n}(x+\rho)=\left\{\begin{array}{ll}\sqrt{\pi} 2^{n} n !(2\sigma)^{m-n} L_n^{m-n}(-2\sigma \rho) & n<m \\ \sqrt{\pi} 2^{m} m !(2 \rho)^{n-m} L_m^{n-m}(-2\sigma \rho) & m<n\end{array}\right\}
\end{align}

\begin{align}\label{eqn:PIElements}
\bra{n}\hat{\Pi}(p, q)\ket{m}=\left\{\begin{array}{ll}
2(-1)^n \sqrt{\frac{2^{m} n!}{2^{n} m !}}e^{-|\alpha|^2}\alpha^{m-n}L_n^{m-n}(2|\alpha|^2) & n<m\\
2(-1)^m \sqrt{\frac{2^{n} m!}{2^{m} n !}}e^{-|\alpha|^2}\alpha^{*n-m}L_m^{n-m}(2|\alpha|^2) & m<n
\end{array}\right\},
\end{align}
where $\alpha := q+ip$. 
 
For a general beamsplitter, the measured overlap is between the Wigner function of the unknown state and a Wigner function of the form given as
\begin{align}\label{eqn:WigFunct}
W_T(\alpha)=\frac{1}{\pi\sigma^2}\exp\left\{-\frac{|\alpha-\beta|^2}{\sigma^2}\right\}.
\end{align}
Defining $\tau=1/\sigma$ and using Eq.~(\ref{eqn:PIElements}), we have that the matrix elements Eq.~(\ref{eqn:OpEle}) of the operator given by the Wigner function above are for $m<n$:
\begin{align}
	T_{n,m}&=\int_{-\infty}^{\infty}2(-1)^m \sqrt{\frac{2^{n} m!}{2^{m} n !}}e^{-|\alpha|^2}\alpha^{*n-m}L_m^{n-m}(2|\alpha|^2)\frac{1}{\pi\sigma^2}\exp\left\{-\frac{|\alpha-\beta|^2}{\sigma^2}\right\}d^2\alpha \nonumber\\
	&=Ce^{-\tau^2|\beta|^2}\int_{-\infty}^{\infty}\alpha^{*n-m}L_m^{n-m}(2|\alpha|^2)e^{-(\tau^2+1)|\alpha|^2}e^{\tau^2(\alpha\beta^*+\alpha^*\beta)}d^2\alpha,
	\label{eq:Polar_trans}
 \end{align}
where we define $C=\frac{2(-1)^m}{\pi\sigma^2}\sqrt{\frac{2^{n} m!}{2^{m} n !}}$. Now we we transform Eq.~\eqref{eq:Polar_trans} using polar coordinate transformation $\alpha=re^{i\theta}$ and $d^2\alpha=rdrd\theta$ which leads to the matrix elements
\begin{align}
	{T}_{n,m}&=Ce^{-\tau^2|\beta|^2}\int_{-\infty}^{\infty} dr \int_{0}^{2\pi} d \theta r^{n-m+1}	e^{-i(n-m) \theta} e^{-(\tau^2+1) r^{2}} e^{\tau^2r\left(e^{i \theta} \beta^{*}+e^{-i \theta}  \beta\right)} L_n^{n-m}\left(2 r^{2}\right)\nonumber\\
	&=Ce^{-\tau^2|\beta|^2}\int_{-\infty}^{\infty}  dr \int_{0}^{2\pi} d \theta r^{n-m+1}	e^{-i(n-m) \theta} e^{-(\tau^2+1) r^{2}}  L_n^{n-m}\left(2 r^{2}\right) \sum_{k=0}^{\infty}  \frac{\tau^{2k}r^{k}}{k!}\left(e^{i \theta} \beta^{*}+e^{-i \theta} \beta\right)^{k}\nonumber\\
	&=Ce^{-\tau^2|\beta|^2}\int_{-\infty}^{\infty}  dr  r^{n-m+1} e^{-(\tau^2+1) r^{2}}  L_n^{n-m}\left(2 r^{2}\right) \sum_{k=0}^{\infty}  \frac{\tau^{2k}r^{k}}{k!}\sum_{l=0}^{k}\beta^{*l}\beta^{k-l}\binom{k}{l} \int_{0}^{2\pi} d \theta e^{i \theta(2l-k-n+m)}.
	\end{align}
	The last integral is null for $l\neq \frac{1}{2}(k+n-m)$ and equals $2\pi$ for  $l= \frac{1}{2}(k+n-m)$. Therefore, we can write $k=n-m+2s$ ($k+n-m$ must be even and $0\leq l\leq k$), $s=0,1,...$ which implies that $l=n-m+s$. This simplification leads to
\begin{align}
	{T}_{n,m}&= 2 \pi C e^{-\tau^2|\beta|^2} \sum_{s=0}^{\infty}\binom{n-m+2 s}{s+n-m}\frac{\beta^{*s+n-m} \beta^{s}}{(n-m+2 s) !}\tau^{2(n-m+2 s)}\nonumber\\
	&\qquad\qquad\qquad\qquad\times\int_{-\infty}^{\infty} d r \, r^{2(n-m+s)+1} e^{-(\tau^2+1) r^{2}} L_{n}^{n-m}\left(2 r^{2}\right) e^{-s r^{2}}. 
\end{align}
To evaluate the last integral (which we will called $I$), we use the following identity \cite{Poh-aun2001}:
\begin{gather}
\int_{0}^{\infty} x^{\mu-1} e^{-\sigma x} L_{n_{1}}^{\left(\alpha_{1}\right)}\left(\lambda_{1} x\right) \cdots L_{n_{r}}^{\left(\alpha_{\nu}\right)}\left(\lambda_{\nu} x\right) d x\overset{x=r^2}{=}2 \int_{0}^{\infty}  r^{2 \mu-1} e^{-\sigma r^{2}} L_{n_{1}}^{(\alpha_1)}\left(\lambda_{1} r^{2}\right) \cdots L^{\alpha_{\nu}}_{n_\nu} \left(\lambda_{\nu} r^{2}\right)d r\nonumber\\
=\left(\begin{array}{c}
n_{1}+\alpha_{1} \\
n_{1}
\end{array}\right) \ldots\left(\begin{array}{c}
n_{\nu}+\alpha_{\nu} \\
n_{\nu}
\end{array}\right) \frac{\Gamma(\mu)}{\sigma^{\mu}} F_{A}^{(r)}\left[\mu,-n_{1}, \ldots,-n_{\nu} ; \alpha_{1}+1, \ldots, \alpha_{\nu}+1 ; \frac{\lambda_{1}}{\sigma}, \ldots, \frac{\lambda_{\nu}}{\sigma}\right] \\
\left(Re(\mu)>0 ; \quad Re(\sigma)>0 ; \quad n_{j} \in \mathbb{N}_{0} ; \quad j=1, \ldots, \nu\right),\nonumber
\end{gather}
where $F_{A}^{(\nu)}$ denotes the first of the four Lauricella's hypergeometric functions of $\nu$ variables defined by 

\begin{align}
F_{A}^{(\nu)}\left[a, b_{1}, \ldots, b_{\nu} ; c_{1}, \ldots, c_{\nu} ; z_{1}, \ldots, z_{\nu}\right]=\sum_{k_{1}, \ldots, k_{\nu}=0}^{\infty} \frac{(a)_{k_{1}+\cdots+k_{\nu}}\left(b_{1}\right)_{k_{1}} \cdots\left(b_{\nu}\right)_{k_{\nu}}}{\left(c_{1}\right)_{k_{1}} \cdots\left(c_{\nu}\right)_{k_{\nu}}} \frac{z_{1}^{k_{1}}}{k_{1} !} \cdots \frac{z_{\nu}^{k_{\nu}}}{k_{\nu} !} \nonumber\\
\left(\left|z_{1}\right|+\cdots+\left|z_{\nu}\right|<1\right)\quad\text{and }(a)_n=\frac{\Gamma(a+n)}{\Gamma(a)},
\end{align}
where $\Gamma(a) = (a-1)!$ are standard gamma functions. Thus,
\begin{align}
I=\frac{1}{2}\frac{n!}{(\tau^2+1)^{n-m+s+1}}\sum_{k=0}^{m}\frac{(-1)^k(n-m+s+k)!}{(m-k)!(n-m+k)!k!}\left(\frac{2}{\tau^2+1}\right)^k,
\end{align}
and the matrix elements to be calculated take the form
\begin{align}
	{T}_{n,m}&= 2 \pi C e^{-\tau^2|\beta|^2} \beta^{*(n-m)} \frac{\tau^{2(n-m)}}{\left(1+\tau^{2}\right)^{n-m+1}} \nonumber\\
	&\qquad\times\sum_{s=0}^{\infty}\binom{n-m+2 s}{n-m+s} \frac{\left(\tau^{4}|\beta|^{2}\right)^{s}n!}{(n-m+2s)!(1+\tau^{2})^s} \sum_{k=0}^{m}\frac{(-1)^{k}(n-m+s+k)!}{(m-k)!(n-m+k)!k!}\left(\frac{2}{\tau^2+1}\right)^k,
\end{align}
where the expression above can be rewritten as
\begin{align}
	{T}_{n,m}&= 2 \pi C e^{-\tau^2|\beta|^2}n! \frac{\beta^{*(n-m)}}{1+\tau^{2}} \left(\frac{\tau^{2}}{1+\tau^{2}}\right)^{n-m}\sum_{k=0}^{m}\frac{(-1)^{k}\left(\frac{2}{\tau^2+1}\right)^k}{(m-k)!(n-m+k)!}\nonumber\\
&\qquad\times\sum_{s=0}^{\infty}\left(\frac{\tau^{4}|\beta|^{2}}{1+\tau^{2}}\right)^{s}\frac{1}{s!} \binom{n-m+s+k}{k}.
\end{align}
Using Vandermonde's identity, we may write the last binomial term as
\begin{align}
	\binom{n-m+s+k}{k}=\sum_{j=0}^{k}\binom{s}{j}\binom{n-m+k}{k-j},
\end{align}
which gives us
\begin{align}
{T}_{n,m}&= 2 \pi C e^{-\tau^2|\beta|^2} \frac{\beta^{*(n-m)}}{1+\tau^{2}} \left(\frac{\tau^{2}}{1+\tau^{2}}\right)^{n-m}\sum_{k=0}^{m}\binom{n}{m-k}\left(\frac{-2}{\tau^2+1}\right)^k \sum_{j=0}^{k}\binom{n-m+k}{k-j}\nonumber\\
&\qquad	\times\sum_{s=0}^{\infty}\left(\frac{\tau^{4}|\beta|^{2}}{1+\tau^{2}}\right)^{s}\frac{1}{s!}\binom{s}{j}\\
&= 2 \pi C e^{-\tau^2|\beta|^2} \frac{\beta^{*(n-m)}}{1+\tau^{2}} \left(\frac{\tau^{2}}{1+\tau^{2}}\right)^{n-m}\sum_{k=0}^{m}\binom{n}{m-k}\left(\frac{-2}{\tau^2+1}\right)^k\nonumber\\
&\qquad	\times\sum_{j=0}^{k}\binom{n-m+k}{k-j}\left(\frac{\tau^{4}|\beta|^{2}}{1+\tau^{2}}\right)^{j}\frac{1}{j!}\exp\left(\frac{\tau^{4}|\beta|^{2}}{1+\tau^{2}}\right)\\
&= 2 \pi C \exp\left(-\frac{\tau^{2}|\beta|^{2}}{1+\tau^{2}}\right) \frac{\beta^{*(n-m)}}{1+\tau^{2}} \left(\frac{\tau^{2}}{1+\tau^{2}}\right)^{n-m}\sum_{k=0}^{m}\binom{n}{m-k}\left(\frac{-2}{\tau^2+1}\right)^kL_k^{n-m}\left(\frac{\tau^{4}|\beta|^{2}}{1+\tau^{2}}\right),
\end{align}
where we have used the additional identities
\begin{align}
	\sum_{s=0}^{\infty}\frac{x^s}{s!}\binom{s}{j}=\sum_{s=j}^{\infty}\frac{x^s}{s!}\binom{s}{j}=\frac{x^je^x}{j!}
\end{align}and the definitions of the generalized Laguerre's polynomials
\begin{align}
\sum_{j=0}^{k}\binom{n-m+k}{k-j}\frac{x^j}{j!}=L_k^{n-m}(-x).
\end{align}
Finally, we can use the multiplication theorem of the generalized Laguerre's polynomials, 
\begin{align}
L_{m}^{\lambda}(y x)=\sum_{k=0}^{m}\binom{m+\lambda}{m-k} L_{k}^{\lambda}(y) x^{k}(1-x)^{m-k},
\end{align}
written as 
\begin{align}
(1-x)^mL_{m}^{n-m}\left(\frac{-yx}{1-x}\right)=\sum_{k=0}^{m}\binom{n}{m-k} L_{k}^{n-m}(y) (-x)^{k},
\end{align}
to derive a closed form for the photon-number basis matrix elements of the operator described by the general Gaussian Wigner function (\ref{eqn:WigFunct}) as
\begin{align}\label{eqn:TmatrixEl}
{T}_{n,m}=\frac{2(-1)^{m}}{2+\tau^2} \sqrt{\frac{2^{n} m !}{2^{m} n !}} \beta^{*(n-m)}\left(\frac{\tau^2}{1+\tau^2}\right)^{n-m}\left(\frac{\tau^{2}-1}{\tau^{2}+1}\right)^{m} L_m^{n-m}\left(\frac{2 \tau^{4}|\beta|^{2}}{\tau^4-1}\right).
\end{align}
This expression allows us to explicitly write down the overlap integral, even in the case of unbalanced beamsplitter, as
\begin{align}
 \mathcal{O}=\sum_{n=0}\sum_{m=0}{T}_{n,m}\rho_{m,n},
\end{align}
where ${T}_{n,m}$'s are  calculated in Eq.~(\ref{eqn:TmatrixEl}) and $\rho_{m,n}$'s are matrix elements of the unknown state.

\section{Tomographic reconstruction with SDP}\label{sec:tomo}
We now use the formalism discussed above to perform the complete state tomography of an arbitrary quantum state. For a given single-mode quantum state, one can write the density matrix in the photon-number basis as
\begin{equation}
    \rho = \sum_{n,n'=0}^{\infty}\rho_{nn'}|n\rangle \langle n'|. 
\end{equation}
 Complete characterization of $\rho$ requires determining $\rho_{nn'}$for all $n, m$.  To do that, we choose a set of distinct coherent states, $|{\alpha_j}\rangle$. Using Eq.~\eqref{eq:meas_overlap}, we obtain the fidelity between $|{\alpha_j}\rangle$ and $\rho$, formulated as
\begin{equation}
    \mathcal{O}^j = Tr[|{\alpha_j}\rangle \langle \alpha_j|\rho] = \langle {\alpha_j}| \rho |\alpha_j\rangle.
\label{eq:fidj}
\end{equation}
Using Eq.~\eqref{eq:fidj} and the coherent state represented in the photon-number basis, $|\alpha_j\rangle = \sum_{n=0}^{\infty}c_{jn}|n\rangle$,  we get
\begin{equation}
    \mathcal{O}^j = \sum_{m'=0}^{\infty}c^{*}_{jm'}\langle m'|\sum_{n,n'=0}^{\infty}\rho_{nn'}|n\rangle \langle n'| \sum_{n'=0}^{\infty}c_{jm}|m\rangle.
\end{equation}
Further simplification leads to
\begin{equation}
     \mathcal{O}^j =\sum_{n,m =0}^{\infty}c_{jm} c^{*}_{jn} \rho_{nm}. 
\end{equation}
Ideally, the sum over $n,m$ goes to infinity but for practical purposes one needs to truncate it at, say $n_0$, such that any terms $n,m>n_0$ do not significantly contribute to the sum. As a result, we have
\begin{equation}
     \mathcal{O}^j = \sum_{n,m =0}^{n_0}c_{jm} c^{*}_{jn} \rho_{nm},
     \label{eq:fid_1}
\end{equation}

where $c_{jm} c^{*}_{jn} = e^{-|\alpha_j|^2}\frac{(\alpha_j^{*})^n (\alpha_j)^m}{\sqrt{n!m!}}$.

Furthermore, by using $N_p = (n_0+1)^2$ coherent states, Eq.~\eqref{eq:fid_1} can be written in the matrix form as 
\begin{equation}
\begin{pmatrix}
\mathcal{O}^{(0)} \\
\mathcal{O}^{(1)} \\
\vdots\\
\mathcal{O}^{N_p}
\end{pmatrix}=\begin{pmatrix}
c^0_0c_0^{0*}& c_0^0c_1^{0*}&\hdots&c_{n_0}^0 c_{n_0}^{0*} \\
c_0^{1}c_0^{1*}& c_0^{1}c_1^{1*}&\hdots&c_{n_0}^{1}c_{n_0}^{1*} \\
\vdots&\vdots&\ddots&\vdots\\
c_{0}^{N_p}c_0^{N_p*}& c_{0}^{N_p}c_1^{N_p*}&\hdots&c_{n_0} ^{N_p}c_{n_0}^{N_p*} \\
\end{pmatrix}
\begin{pmatrix}
\rho_{0,0} \\
\rho_{0,1}\\
 \vdots \\
 \rho_{n_0,n_0}
\end{pmatrix}.
\label{eq:matrix}\nonumber\\
\end{equation}
We can rewrite the above matrix equation in compact form as
\begin{equation}\label{eq:compact}
    \bf{O} = {\bf{CP}},
\end{equation}
where ${\bf{O}}\in \mathbb{R}^{(n_0+1)^2}$, ${\bf{P}}\in \mathbb{C}^{(n_0+1)^2}$ and ${\bf{C}}\in \mathbb{C}^{(n_0+1)^2\times(n_0+1)^2}$. 
Next, we can invert Eq.~\eqref{eq:compact} to reconstruct ${\bf{P}}$. This can be achieved by solving the following semidefinite program. 
\begin{align}
\underset{\bf{P}}{\text{Minimize}}\quad\quad&||\bf{O-CP}||_2\nonumber\\
\text{Subject to}\quad&\rho\geq0\quad \text{and}\quad \text{Tr}[\rho] = 1,
\end{align}
where $||.||_2$ is the $l_2$ norm defined as $||V||_2 = \sqrt{\sum_{i}|v_i|^2}$. Note that this kind of quadratic convex techniques have been extensively discussed in the the context of quantum detector tomography~\cite{Lundeen2009NP,Grandi_2017,Zhang_2012,nehra2019characterizing}.  The optimization problem is convex and can be efficiently solved for a guaranteed unique ${\bf{P}}$, and hence for the unknown state, $\rho$, using open source Python module CVXPY~\cite{cvxpy}. Although this method holds for general quantum states, we restrict our simulations to real-valued density matrices for numerical ease. However, we do show the reconstruction for complex-valued density matrices in Sec.~\ref{sec:complex}.

All numerical simulations were performed in QuTip~\cite{johansson2013qutip} where the Hilbert space for each optical mode was constructed in the Fock basis with a high enough dimensionality to ensure state probability amplitudes decayed to less than $10^{-7}$ before truncation. Under these parameters, the numerically efficient SDP algorithms converged on the order of $10^{-2}$ seconds 
on a 3GHz Intel i5 quad core processor with 16Gb RAM.\\

Our method is demonstrated in Fig. \ref{fig:dm_reconstruction} for the example cases of the small amplitude cat state, $|\psi\rangle \propto |\alpha\rangle+|-\alpha\rangle$ where $\alpha=\sqrt{3}$, and a Gottesman-Kitaev-Preskill (GKP) state of mean photon number 5.  These states were reconstructed using 400 different probing coherent states of 20 amplitude increments from $\beta=0$ to $\beta=\sqrt{6}$ and 20 phase increments from 0 to $2\pi$, to achieve fidelities with the target states greater than $0.999$ for the cat state, and a fidelity of $0.985$ for the GKP state. Because the observer is assumed to have no prior knowledge of the state to be characterized, it is important to scan the entirety of phase space in question with different coherent states so as to have sufficient overlap between all portions of the state under test.  If some prior knowledge of the state is obtained, then the probing coherent states can be restricted to a localized region of phase space near the unknown quantum state.\\
\begin{figure}[!h]

    \includegraphics[width = 0.85\textwidth]{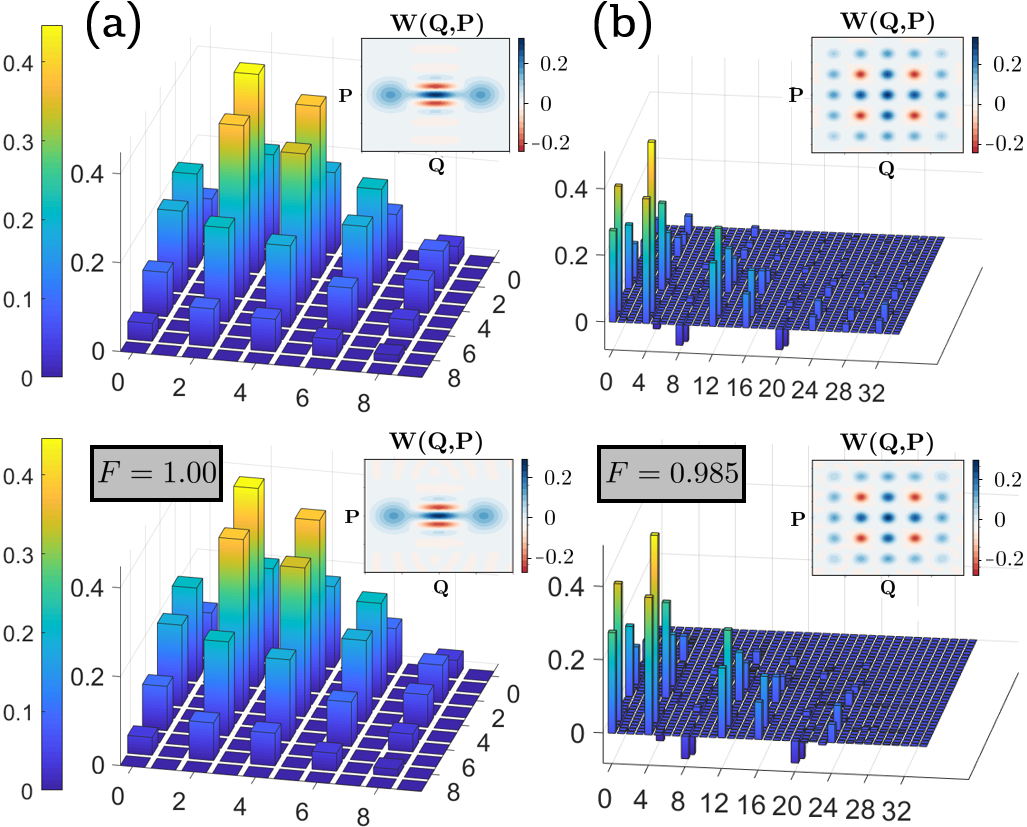}
    \caption{Tomographic reconstruction using 400 coherent state probes for  \textbf{(a)} a cat state of amplitude $\sqrt{3}$, and \textbf{(b)} a GKP state with a mean photon number of 5. The top row displays the  density matrix for the ideal theoretical state, and the bottom shows the reconstructions. Insets display the plotted Wigner function for each state.} 
    \label{fig:dm_reconstruction}
\end{figure}

\subsection{Complex Reconstruction}\label{sec:complex}
We demonstrate the tomography protocol for complex-valued density matrices displayed in Fig.~\ref{fig:complex_rho}.  We perform the tomography with coherent state probes that range in amplitude in 20 steps from $|\beta| \in [0,\sqrt{3} ]$ and 20 phases $\phi \in [0, 2\pi]$, for a coherent state denoted by complex variable, $\alpha = \sqrt{2}(i+1)$ and a superposition of photon-number states with the complex probability amplitude.  The Wigner functions are shown along with separate plots for the real and imaginary elements of the respective reconstructed density matrices, including an inset fidelity with the ideal states.
\begin{figure*}[!hbt]
    \centering

\includegraphics[width = 0.9\textwidth]{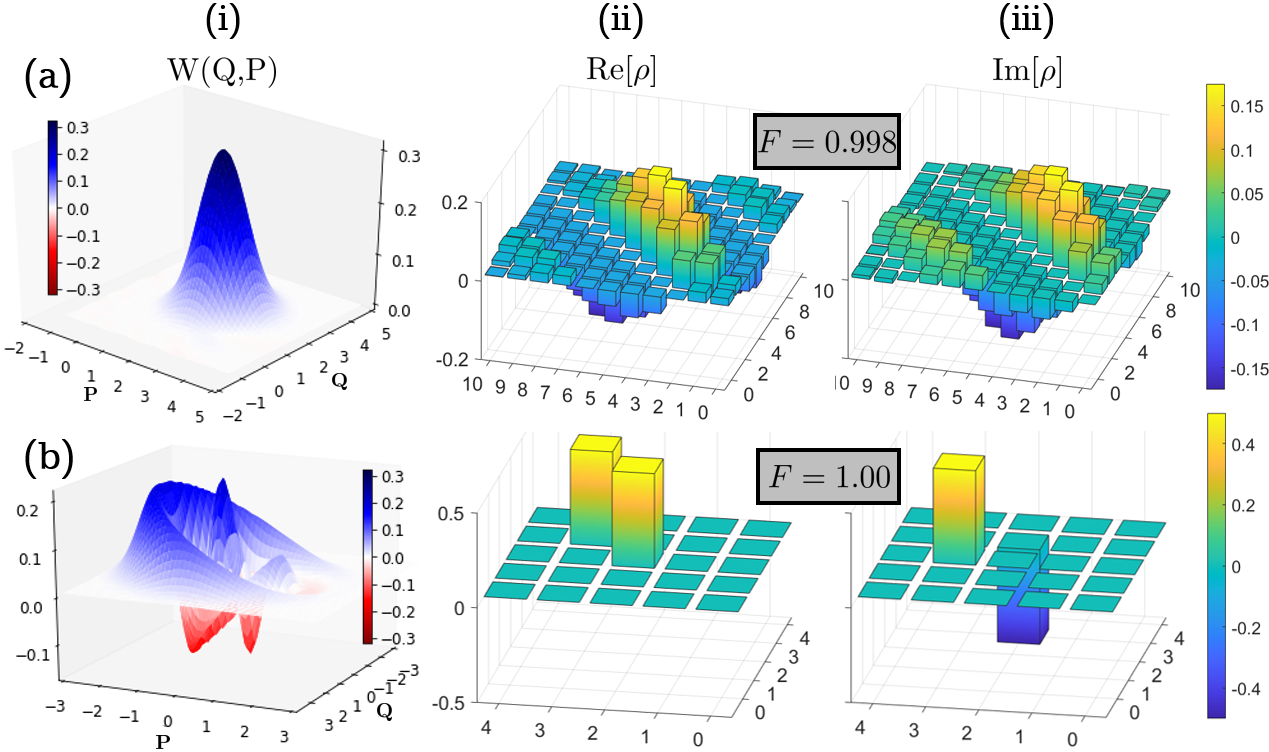}
\caption{Reconstruction of states with complex-valued density matrices for (a) the coherent state, $|\alpha\rangle$, with $\alpha=\sqrt{2}(i+1)$ and (b) the superposition $\frac{1}{\sqrt{2}}(|2\rangle - i|3\rangle)$. The inset fidelity is calculated between the reconstructed state and the ideal target state. (i) Reconstructed Wigner functions. (ii) Real elements of the density matrices. (iii) Imaginary elements of the density matrices. }
\label{fig:complex_rho}
\end{figure*}

\section{Experimental set-up}\label{sec:exp}
The tomography of a quantum states is performed by interfering a mode-matched local oscillator (LO) with the signal state, $\rho$, at a balanced beamsplitter followed by detection of one output mode using a photon-number resolving transition-edge sensor (TES) as shown in Fig.~\ref{fig:exp_setup}. A portion of the LO is split and strongly attenuated by neutral density (ND) filters to be used as a coherent state, $|\beta\rangle$, for the signal when the flip mirror is engaged. When the flip mirror is not in place, the signal is a single-photon source based on heralded, cavity-enhanced type-II spontaneous-parametric downconversion from a periodically poled potassium titanyl phosphate (ppKTP) 
crystal. Spectral and spatial filtering was achieved by the optical parametric oscillator created by placing the crystal in a resonant cavity and an additional Fabry-Perot filter cavity on the heralding arm as shown in Fig.~\ref{fig:exp_setup}. The cavities were Pound-Drever-Hall-locked~\cite{Drever1983} using a portion of the LO in an ``on/off" configuration as described in Ref.~\cite{Nehra2019}. The coherent-state probes derived from the LO were amplitude modulated with a combination of polarizer and electro-optic modulator (EOM) and were phase controlled with a mirror-mounted piezoelectric actuator (PZT). Imperfections in phase control and stability resulted in approximately $0.05$ radians of phase-error on probe calibrations that contribute to the asymmetric ripples seen in the experimentally constructed Wigner functions in the main text. Extensive details on the single-photon source, mode filtering, and the LO amplitude calibration using the TES can also be found in Ref.~\cite{Nehra2019}.  
\begin{figure}[h]
    \centering

\includegraphics[width = 0.99\textwidth]{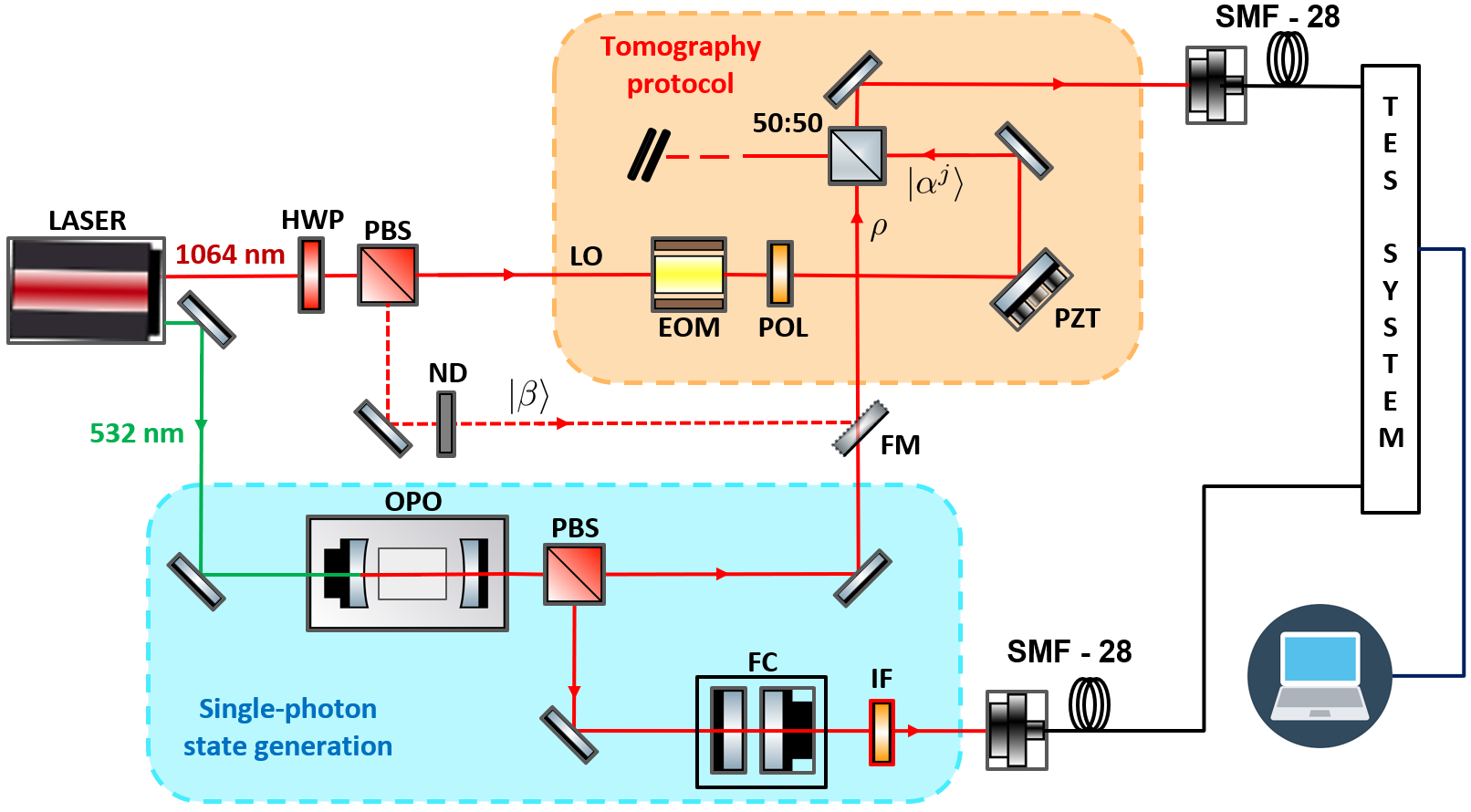}
\caption{Experimental setup. The tomography protocol is contained in the pink box where the mode-matched LO is interfered with the state $\rho$. When the FM is in position, tomography for the coherent state, $\beta$, is performed; otherwise, $\rho$ is the single-photon state generated in the blue box.  EOM, electro-optic modulator; FC, filter cavity; FM, flip mirror; HWP, half-wave plate; IF, interference filter; LO, local oscillator; ND, neutral-density filter; OPO, optical parametric oscillator; PBS, polarizing beamsplitter; POL, polarizer; PZT, piezoelectric transducer.}
    \label{fig:exp_setup}
\end{figure}
\section{Loss Compensation}\label{sec:loss_comp}
\subsection{Complete density matrix correction}\label{sec:density_correction}
We now wish to correct an arbitrary density matrix given a known loss.  In this case, we have experimentally measured $\rho'$, but our goal is to reconstruct the density matrix before the loss, $\rho$.  As shown in Fig. \ref{fig:loss_bs}, this can be modeled by sending $\rho$ through a fictitious `loss beamsplitter' with reflection and transmission coefficients of $r=\sqrt{1-\eta}$ and $t=\sqrt{\eta}$, where $\eta$ is the overall transmission efficiency.
\begin{figure}[h]
    \centering

\includegraphics[width = 0.35\textwidth]{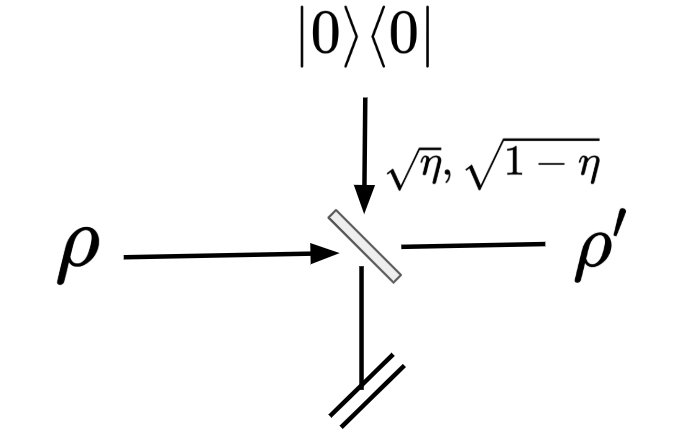}
\caption{Lossy channel.}
    \label{fig:loss_bs}
\end{figure}

The general quantum state density matrix before the loss can be written as
\begin{align}
    \rho &= \sum_{n,n' =0}^{\infty}\rho_{n,n'}|n\rangle\langle n'| \nonumber \\
    \rho &=\sum_{n,n' =0}^{\infty}\frac{\rho_{n,n'}}{\sqrt{n!n'!}}{a^\dag}^n|0\rangle\langle 0|a^{n'}.
\end{align}
If this state enters into the loss beamsplitter in mode $\hat{a}$ with vacuum in mode $\hat{b}$, then the mode operators transform in the Heisenberg picture according to $\hat{a} \rightarrow t\hat{a} + r\hat{b}$ and  $\hat{b} \rightarrow -r\hat{a} + t\hat{b}$ to yield an output density matrix
\begin{equation}
    \rho_{out} = \sum_{n,n' =0}^{\infty}\rho_{n,n'}\frac{(t\hat{a}^\dag+r\hat{b}^\dag)^n}{\sqrt{n!}}|0\rangle_{a}|0\rangle_{b} \!\langle 0|{}_b\!\langle 0|_a\frac{(t\hat{a}+r\hat{b})^{n'}}{\sqrt{n'!}}.
\end{equation}
Tracing out over mode $b$ yields the final state after loss, which is given by
\begin{equation}
\begin{aligned}
   \rho'= Tr_b[\rho_{out}]  =\sum_{n,n' =0}^{\infty}\rho_{n,n'}\sum_{k =0}^{n}\sum_{k'=0}^{n'}A_{n,n',k,k'}|n-k\rangle \langle{n'-k'}|\langle k|k'\rangle\delta_{k,k'},
\end{aligned}
\end{equation}
where
\begin{equation}
    A(n,n',k,k')=\sqrt{{\binom{n}{k}} {{n'}\choose{k'}}}r^{k+k'}t^{n+n'-k-k'}.
\end{equation}
Substituting $n-k$ and $n'-k$ with $m$ and $m'$ allows us to rearrange the expression and write a sum over the Fock components in order, which can be written as

\begin{equation}
    \rho'=\sum_{m,m',k =0}^{\infty}\rho_{(m+k),(m'+k)}A(m+k,m'+k,k,k)|m\rangle \langle{m'}|,
\end{equation}
where it is easy to see that each element of the density matrix after loss is related to the original state by
\begin{equation}
    \rho_{m,m'}'=\sum_{k =0}^{\infty}\rho_{(m+k),(m'+k)}\sqrt{{{m+k}\choose{k}} {{m'+k}\choose{k}}}r^{2k}t^{m+m'}.
    \label{eq:rho_loss_11}
\end{equation}
This can be viewed as a generalized Bernoulli distribution~\cite{Kiss1995}, so can be inverted to read
\begin{equation}
    \rho_{m,m'}=\sum_{k =0}^{\infty}\rho_{(m+k),(m'+k)}'\sqrt{{{m+k}\choose{k}} {{m'+k}\choose{k}}}(-1)^k\Big(\frac r t\Big)^{2k}t^{-m-m'}.
    \label{eq:rho_loss_supp}
\end{equation}
  In practice, the sum over $k$ can be truncated to some value, $N_{max}$, beyond which the entries in the initial density matrix are negligible.  We can then reformulate Eq. \ref{eq:rho_loss_11} as a series of $N_{max}$ linear maps from the $i^{th}$ diagonal of $\rho'$ to the $i^{th}$ diagonal of $\rho$, where the main diagonal is given by $i=0$. Each of these linear maps, $\textbf{M}^{(i)}$, is an upper triangular matrix of dimension $N_{max}-i \times N_{max}-i$ with elements
\begin{equation}
 {\textbf{M}_{jk}^{(i)}}(\eta) = \Bigg\{ \begin{array}{cc} 
               0 & \hspace{5mm} j > k \\
                \sqrt{{{k}\choose{k-j}} {{i+k}\choose{k-j}}}(1-\eta)^{(k-j)}\eta^{\frac i2+j} & \hspace{5mm} \text{otherwise} \\
                \end{array} 
\end{equation}

Since each $\mathbf{M}^{(i)}(\eta)$ is triangular with nonzero diagonal elements, the inverse mappings can be found by inverting the generalized Bernoulli transformation and are given by~\cite{Kiss1995}
\begin{align}
  \text{Inv}[{\mathbf{M}^{(i)}(\eta)}]=\mathbf{M}^{(i)}(\eta^{-1}).
\label{eq:M_inverse}
\end{align}
The existence of this inversion is due to the known well defined statistical nature of loss channel, which makes it possible to perfectly reconstruct any $\rho$ within a finite-dimensional Hilbert space when $\eta$ and $\rho'$ are precisely known~\cite{Kiss1995}.  However, the presence of any small deviations in an experimentally measured $\rho'$ can lead to unphysically large or non-positive diagonal density matrix elements in the reconstruction of $\rho$, even while $\rho$ remains normalized, which is similar to the possible numerical instabilities that arise when using pattern-functions~\cite{lvovsky2009continuous}. 
These errors become pronounced for low detector efficiencies at high photon-numbers as seen Fig.~\ref{fig:photon_num_dist}a for the specific case of a loss-compensated cat state. Therefore, it becomes extremely crucial to have \textit{a priori} information about the energy of the quantum states under consideration.

\begin{figure}[ht]
    \centering

\includegraphics[width = 0.9\textwidth]{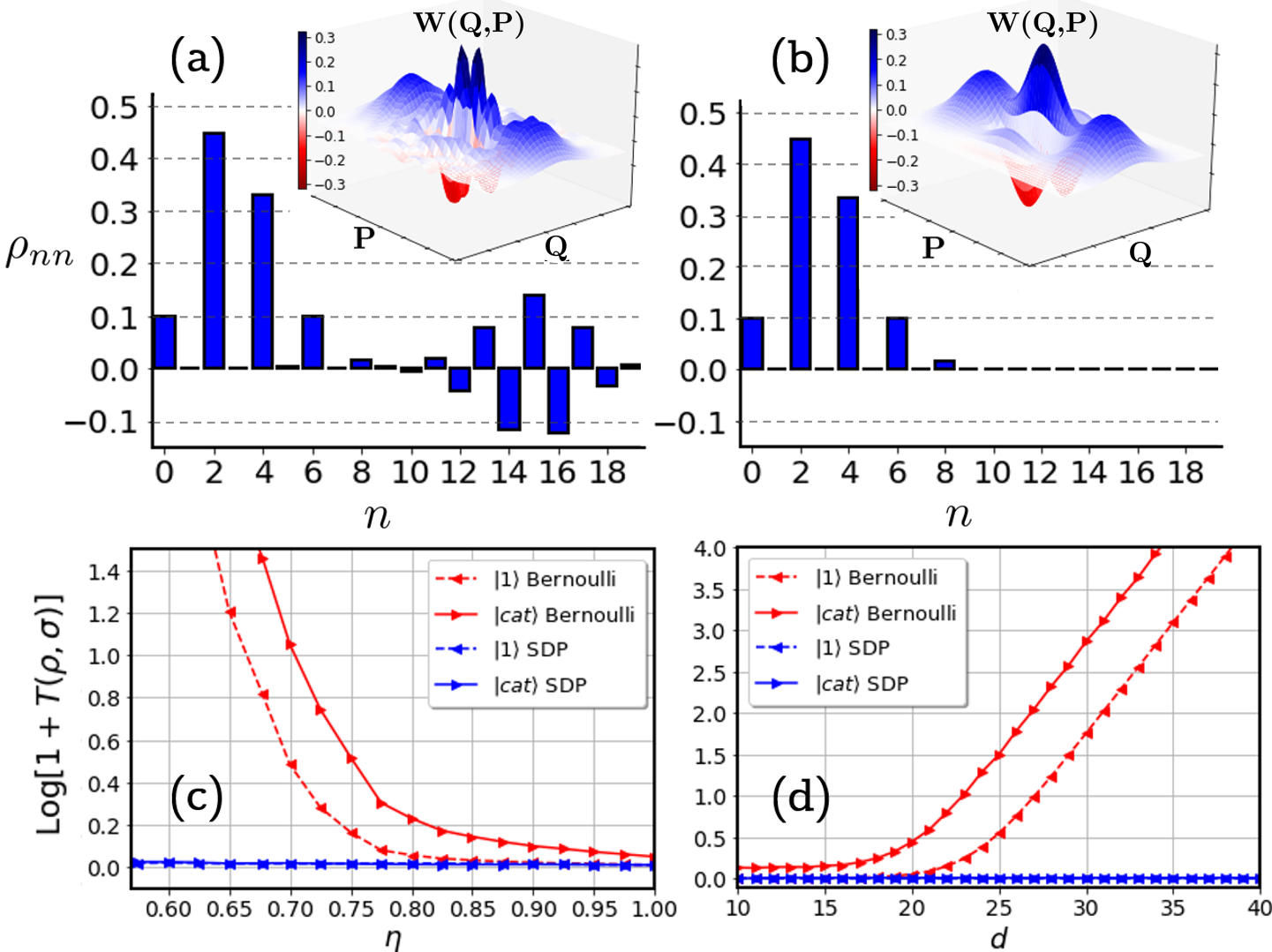}
\caption{Loss-compensation for tomographed cat state of amplitude $\sqrt{3}$ after transmission of $\eta=0.70$ and Hilbert space cut-off of $d=20$, with (a) inversion using the generalized Bernoulli transformation and (b) inversion using SDP. The logarithm of the trace-distance between the reconstructed state, $\rho$, and the target state, $\sigma$, is plotted against $\eta$ [$d$] in (c) [(d)]. We note that $T(\rho,\sigma)>1$ occurs due to the unphysical  reconstruction of $\rho$ and large non-positive diagonal elements. The figure insets show the Wigner function for each state.}
    \label{fig:photon_num_dist}
\end{figure}

Here, we are able to relax this issue by inverting each $\mathbf{M}^{(i)}(\eta)$ using semidefinite programming, where the optimization problem is defined as
\begin{align}
\underset{{\rho}}{\text{Minimize}}\quad&\sum_{i=0}^{N_{max}}||{ \rho'^{(i)}_{diag}-\mathbf{M}^{(i)}\rho^{(i)}_{diag}}||_2\\
\text{Subject to}\quad&\rho\geq0, \hspace{2mm} \text{Tr}[\rho] = 1, \text{ and} \hspace{2mm} \rho_{m,m}\leq \eta^{-m}\rho'_{m,m},\nonumber
\label{eq:minimize_loss}
\end{align}
where $\rho^{(i)}_{diag}$ denotes the $i^{th}$ diagonal of $\rho$ and the third constraint is obtained by noting that each element in the sum in Eq.~\eqref{eq:rho_loss_supp} is positive for $m=m'$, leading to the inequality when the sum is truncated after the first term. Additionally, it is only necessary to sum over the upper diagonals of $\rho$ in the minimization (hence the sum starting at $i=0$), due to the enforced hermiticity of $\rho$.\\
The application of these constraints enforces physicality and avoids the numerically unstable reconstruction that would result by using an exact expression for $ {\mathbf{M}^{(i)}(\eta)}^{-1}$. We demonstrate in Fig.~\ref{fig:photon_num_dist} how small errors on density matrix elements from performing the tomographic procedure on a loss degraded cat state and single-photon Fock state give rise to an unphysical loss-compensated state using the analytical matrix inversion from Ref.~\cite{Kiss1995}, whereas inversion using SDPs successfully reconstructs the state prior to loss. Although all errors in the tomographed density matrix elements prior to loss compensation are on the order of $10^{-3}$ (not depicted), the analytical matrix inversion drastically magnifies these slight deviations. In particular, Fig.~\ref{fig:photon_num_dist}c and Fig.~\ref{fig:photon_num_dist}d show that the validity of the loss-compensation can heavily depend on both the overall loss and on the choice of Hilbert dimension cutoff. When comparing the reconstructed state, $\rho$, to the ideal state without loss, $\sigma$, using the trace distance defined by $T(\rho,\sigma)=\frac{1}{2}||\rho-\sigma||_1$, we see that the deviation of $\rho$ from $\sigma$ grows quickly as $\eta$ shrinks and $d$ increases in the case of analytic inversion. However, $T(\rho,\sigma)$ is both small and relatively independent of either $\eta$ or $d$ when using SDP. As a result, our method is significantly more robust to experimental noise.

\section{Equivalence of photon-number distributions}\label{sec:pn_dist_equiv}
To show the equivalence of the photon-number distributions measured in each configuration in Fig.~\ref{fig:loss_model}, we adapt the approach originally introduced in~\cite{Wallentowitz1996}. The signal and LO modes are described by annihilation operators $\hat{a}$ and $\hat{b}$ respectively, and $\hat{c}_v$ and $\hat{d}_v$ are vacuum modes. 
\begin{figure*}[!hbt]
    \centering
	    \includegraphics[width =1\textwidth]{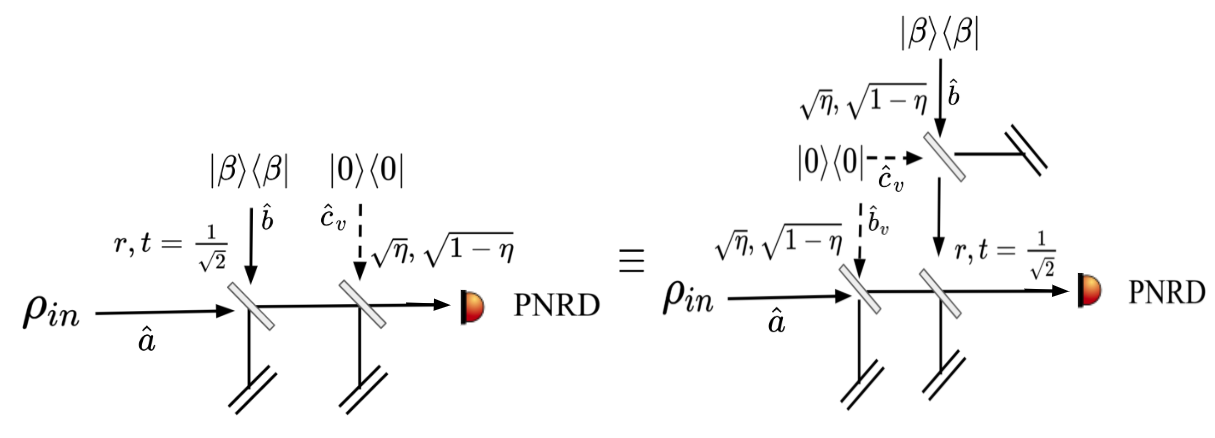}
    \caption{Schematic of the loss model. Left and right networks produce the same photon-number distribution.}
    \label{fig:loss_model}
\end{figure*}
For a perfect PNR detector,  the probability of measuring $n$ photons is given by~\cite{Kelley1964}
\begin{equation}
    P(N=n) = \bigg\langle :\frac{\hat{N}^n}{n!}e^{-\hat{N}}:\bigg\rangle_{\rho_{in}},
    \label{eq:mandel_wolf_formulla}
\end{equation}
where $\hat{N} = \hat{d}^\dagger \hat{d}$ is the photon-number operator of the detection mode and the expectation value is calculated over the initial states, and $::$ is the normal ordering. By employing the Heisenberg picture, we first determine the detection mode in terms of input modes for the network on the left of Fig.~\ref{fig:loss_model}. 
 The input mode denoted by annihilation operator, $\hat{a}$, evolves to 
\begin{align}
    \text{After first beamsplitter:}~~\hat{a}&\rightarrow \frac{\hat{a}+\hat{b}}{\sqrt{2}}\\ \nonumber
   \text{After second beamsplitter:}~~&
   \sqrt{\eta}\bigg(\frac{\hat{a}+\hat{b}}{\sqrt{2}}\bigg) + \sqrt{1-\eta}\hat{c}_v\\
\end{align}
Since the input states for mode $\hat{b}$ and $\hat{c}_v$ are coherent and vacuum states respectively, the normal ordering allows to treat them as  complex numbers. As a result, the effective photon-number operator is given by
\begin{equation}
    \hat{N}^{L}_{\text{eff.}} = \hat{d}^\dagger \hat{d},
\end{equation}
where the detection mode is
\begin{equation}
    \hat{d}^L = \sqrt{\eta}\bigg(\frac{\hat{a}+\beta}{\sqrt{2}}\bigg)\\
    \label{eq:dL}
\end{equation}
Likewise, for the right network, we have
\begin{align}
    \text{After left beamsplitter:}~~\hat{a}&\rightarrow \sqrt{\eta}\hat{a}+\sqrt{1-\eta}\hat{b}_v\\ 
   \text{After top beamsplitter:}~~\hat{b}&\rightarrow \sqrt{\eta}\hat{b}+\sqrt{1-\eta}\hat{c}_v\\ \nonumber
   \text{After balanced beamsplitter:}~~\hat{d}&=\frac{1}{\sqrt{2}}(\sqrt{\eta}\hat{a}+\sqrt{1-\eta}\hat{b}_v+\sqrt{\eta}\hat{b}+\sqrt{1-\eta}\hat{c}_v), \nonumber
\end{align}
where $\hat{c}_v$, $\hat{b}_v$ are vacuum modes and $\hat{b}$ is a coherent state. We again utilize the fact that normal ordering allows coherent states to be represented by a complex number and the vacuum state can also be considered as a coherent state with zero amplitude. Thus, the detection mode can be further simplified as
\begin{equation}
    \hat{d}^R = \sqrt{\eta}\bigg(\frac{\hat{a}+\beta}{\sqrt{2}}\bigg).\label{eq:dR}
\end{equation}
From Eq.~\ref{eq:dL} and Eq.~\ref{eq:dR}, one can see that both networks have the same detection mode, therefore would produce the same photon-number distribution for a given quantum state under investigation. 
{
\section{Mode Mismatch Correction}\label{sec:mode_mismatch}}
In this section, we consider the effects of mode mismatch on the measured photon-number distributions. 
In contrast to balanced HD (BHD), the imperfect modematching between the coherent probes and the signal field cannot simply be treated as losses. This can be understood as follows:  In BHD, the measured photocurrent difference is proportional to only the interference term, i.e, $I_{-} \propto \hat{a}^\dagger \alpha_{LO} + \hat{a}\alpha_{LO}^*$, which implies that only the overlapping portion of the signal field gets amplified by the strong LO and the non-overlapping portion is considered as losses. Here, we show that this will no longer be the case with the proposed method. 
\begin{figure}
    \centering
    \includegraphics[width = 0.45\textwidth]{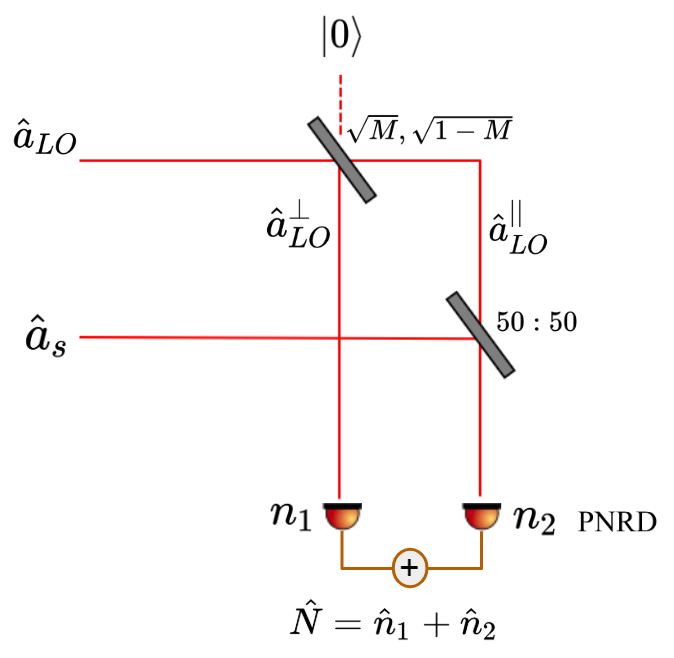}
    \caption{Model for mode mismatch analysis.}
    \label{fig:mode_mismatch_circuit}
\end{figure}
 As displayed in Fig~\ref{fig:mode_mismatch_circuit}, the interference between the local oscillator (LO) and the signal mode, denoted by $\hat{a}_s$ and $\hat{a}_{LO}$, respectively, can be decomposed into two orthogonal modes that each reach the PNR detector. The LO can be seen as interfering with vacuum mode $\hat{a}_{v}$ to be split into a component that overlaps (interferes) entirely with the signal field,  $\hat{a}_{LO}^{||}$, and an orthogonal component,  $\hat{a}_{LO}^{\perp}$, that proceeds to the detector without interacting with the signal. Defining the mode-mismatch parameter, $M$, as the transmission of the fictitious beamsplitter decomposing the components of the LO and making use of the Heisenberg picture, we get

\begin{align}
   \hat{a}_{LO}^{\perp} &=  \sqrt{1-M}\hat{a}_{LO}+\sqrt{M} \hat{a}_{v}.\\
    \hat{a}_{LO}^{||} &=  
    -\sqrt{M}\hat{a}_{LO}+\sqrt{1-M} \hat{a}_{v}.
\end{align}
Likewise, the signal mode after interfering with $\hat{a}_{LO}^{||}$ at the balanced (50:50) beamsplitter evolves to
\begin{equation}
    \hat{a}_s\rightarrow \hat{U}_{BS}\hat{a}_s\hat{U}_{BS}^\dagger =  \frac{\hat{a}_s+\hat{a}_{LO}^{||}}{\sqrt{2}}, 
\end{equation}
where $\hat{U}_{BS}$ is the unitary operator of the balanced beamsplitter.
\begin{align}
    \hat{n}_1 &= (\hat{a}_{LO}^{\perp})^\dagger \hat{a}_{LO}^{\perp}\\
    \hat{n}_2 & = \bigg(\frac{\hat{a}_s+\hat{a}_{LO}^{||}}{\sqrt{2}}\bigg)^\dagger \bigg(\frac{\hat{a}_s+\hat{a}_{LO}^{||}}{\sqrt{2}}\bigg) = \hat{U}_{BS}\hat{a}_s^\dagger\hat{a}_s\hat{U}_{BS}^\dagger
\end{align}
As a result, the total number operator is 
\begin{equation}
    \hat{N} = \hat{n}_1 + \hat{n}_2 = (\hat{a}_{LO}^{\perp})^\dagger \hat{a}_{LO}^{\perp}+ \hat{U}_{BS}\hat{a}_s^\dagger\hat{a}_s\hat{U}_{BS}^\dagger.
\end{equation}
 By employing  Eq.~\ref{eq:mandel_wolf_formulla}, one can further determine the probability of detecting total $n = n_1 + n_2$ photons by both the detectors in Fig.~\ref{fig:mode_mismatch_circuit} as
\begin{equation}
    P(n=n_1+n_2) = \bigg\langle :\frac{\hat{N}^n}{n!}e^{-\hat{N}}:\bigg\rangle_{\rho_{in}}
    \label{eq:mandel_wolf_formulla2},
\end{equation}
where $\hat{N} = \hat{n}_1+\hat{n}_2$ is the two-mode photon-number operator. We then use the fact that in the normal ordering formulation, the annihilation operators denoting coherent states can be simply treated as complex variables, $\alpha_{LO}^\perp$ and $\alpha_{LO}^{||}$. Therefore, we have
\begin{equation}
    \hat{N} = (1-M)|\alpha_{LO}^\perp|^2 + \hat{U}_{BS}\hat{a}_s^\dagger\hat{a}_s\hat{U}_{BS}^\dagger.
    \label{eq:N_simplified}
\end{equation}
Using Eq.~\eqref{eq:mandel_wolf_formulla2} and Eq.~\eqref{eq:N_simplified} results in
\begin{equation}
    P(n) = \bigg\langle:e^{-\big[(1-M)|\alpha_{LO}^\perp|^2 + \hat{U}_{BS}\hat{a}_s^\dagger\hat{a}_s\hat{U}_{BS}^\dagger\big]}\frac{\big[{(1-M)|\alpha_{LO}^\perp|^2 + \hat{U}_{BS}\hat{a}_s^\dagger\hat{a}_s\hat{U}_{BS}^\dagger}\big]^n}{n!}:\bigg\rangle_{\rho_{in}}
\end{equation}
After further simplification, we get
\begin{align}
      P(N=n) &= \bigg\langle:e^{-\big[(1-M)|\alpha_{LO}^\perp|^2 +\hat{U}_{BS}\hat{a}_s^\dagger\hat{a}_s\hat{U}_{BS}^\dagger\big]}
      \sum_{l=0}^{n}{n\choose l}\frac{(\hat{U}_{BS}\hat{a}_s^\dagger\hat{a}_s\hat{U}_{BS}^\dagger\big)^l [(1-M)|\alpha_{LO}^\perp|^2]^{n-l} }{n!}
      :\bigg\rangle_{\rho_{in}}\\
      &=\sum_{l=0}^{n}\bigg\langle:\frac{e^{\hat{U}_{BS}\hat{a}_s^\dagger\hat{a}_s\hat{U}_{BS}^\dagger}(\hat{U}_{BS}\hat{a}_s^\dagger\hat{a}_s\hat{U}_{BS}^\dagger\big)^l}{l!}:\bigg\rangle_{\rho_{in}}
     \frac{e^{-[(1-M)|\alpha_{LO}^\perp|^2]} [(1-M)|\alpha_{LO}^\perp|^2]^{n-l} }{(n-l)!}
     \label{eq:prob_mode_mismatch}
\end{align}
From Eq.~\eqref{eq:prob_mode_mismatch}, one can see that the probability of detecting n photons is the convolution of two probability distributions. The first term in the normal ordering form corresponds to detecting $l$ photons in the signal mode after the interference with $\hat{a}_{LO}^{||}$ while the Poissonian distribution is the probability of $(n-l)$ photons being detected in the orthogonal LO mode, $\hat{a}_{LO}^{\perp}$. We can further rewrite Eq.~\eqref{eq:prob_mode_mismatch} in a compact way as
\begin{equation}
    P(N=n) = \sum_{l=0}^{n}P^{||}(l)P^{\perp}(n-l).
    \label{eq:PNn}
\end{equation}
Note that $P^{\perp}(n-l)$ can be determined by knowing the overlap parameter, $M$, which is experimentally measured from a bright-field visibility measurement~\cite{Banaszek2002}. Next, one can simply invert Eq.~\eqref{eq:PNn} in order to reconstruct the true photon-number distribution in the interfered field of unknown state and the modematched part of LO field, i.e, $|\sqrt{M}\alpha^{||}_{LO}\rangle$. 

\end{widetext}

\end{document}